\documentclass[twocolumn,showpacs,prrprintnumbers,amsmath,amssymb]{revtex4}

\usepackage{amsmath}
\usepackage{graphicx}
\usepackage{dcolumn}
\usepackage{bm}

\begin{document}

\title{Size distribution of emitted energies in local load sharing fiber bundles}
\author{Subhadeep Roy}
\email{subhadeep.roy@ntnu.no} 
\affiliation{PoreLab, Department of Physics, Norwegian University of Science and Technology, NO--7491 Trondheim, Norway.}

\author{Soumyajyoti Biswas}
\email{soumyajyoti.b@srmap.edu.in}
\affiliation{Department of Physics, SRM University - AP, Andhra Pradesh 522502, India.}

\date{\today {}}
\begin{abstract}
\noindent We study the local load sharing fiber bundle model and its energy burst statistics. While it is known that the avalanche size distribution of the model is exponential, we numerically show here that the avalanche size ($s$) and the corresponding energy burst ($E$) in this version of the model have a non-linear relation ($E\sim s^{\gamma}$). Numerical results indicate that $\gamma\approx 2.5$ universally for different failure threshold distributions. With this numerical observation, it is then possible to show that the energy burst distribution is a power law, with a universal exponent value of $-(\gamma+1)$. 
\end{abstract}
\maketitle


\section{Introduction}
It is well known experimentally that quasistatically stressed disordered solids produce intermittent response statistics \cite{phys_rep}, particularly in terms of acoustic emissions, that show scale-free size distributions. These intriguing dynamics is seen universally across scales from microscopic laboratory samples to the geological scale of earthquakes \cite{herrmann90,bkc96,rmp_2012,wiley_book,lucilla}. Empirically, the scale-free size distribution of breaking progression is known in different communities for decades. For example, in geoscience, this is known as the Gutenberg-Richter law, in magnetic domain walls as crackling noise, and so on. \\

The interests of statistical physicists in this context stems from the universal nature of the dynamics across length and energy scales. The scale-free variations of acoustic emissions, waiting time statistics, etc., are independent of the microscopic details of the underlying systems, which are very different from each other. Such behavior indicates critical dynamics, particularly self-organized critical dynamics for the system, where the universality hypothesis is still applicable, without having to fine-tune a driving parameter \cite{bonamy}. Such a phenomenon is therefore open for analysis with the tools of critical phase transitions, universality and therefore is an important step towards predictability of imminent failure \cite{alava_lasse,kun_record,sci_rep}. \\

As a consequence of the scale-free dynamics and potential applicability of the universality hypothesis, many generic models were proposed over the years that reproduce such a scale-free behavior. Such models include the fiber bundle model, the random fuse model, the Burridge-Knopoff model, and so on \cite{wiley_book,fbm_rmp,fbm_book}. The common underlying feature of these models is that they are threshold activated, driven, dynamical models. Particularly, for an external driving parameter crossing a pre-assigned threshold value for a single unit of these models, that unit is activated and influences the units in its `neighborhood’, which may in-turn get activated and thereby initiating an ‘avalanche’. As can be guessed, this type of dynamics is often related to sandpile models of self-organized criticality \cite{soc} and indeed such associations extensively explored in the past \cite{soc_fbm}. \\

The two major parameters that influence the nature of the response in such models are the range of interaction and the strength of the disorder. It was explored, particularly in the fiber bundle model that for a moderate disorder, a scale-free avalanche statistics is only recovered for a `sufficiently’ long-range of interaction \cite{pre1,pre2,kun_ptrc}. In the random fuse model, where the interaction range is not parameters to be tuned, it was shown that the avalanche statistics is not a power-law in the large system size limit \cite{sch}. This is in apparent contradiction with the fact that in reality, the interaction range in disordered elastic samples is not infinite i.e., not a mean-field-like interaction. However, experiments routinely reveal scale-free statistics. \\

One important distinction between the analytical and numerical results of avalanche dynamics and that of the experiments is that in the former it is the number of elements failing in an avalanche that is the measurable quantity, while in the latter it is the energy released in the avalanche. Now, in the mean-field limit of the fiber bundle model, it is straightforward to show that the avalanche size and the energy avalanche size are proportional, hence the two distributions are identical in shape. But this relation is no longer valid for local load sharing variants. In those cases, therefore, it is crucial to explore the size distributions of the energy emissions and compare that with experiments. In this work, we consider the simplest possible variant of the local load sharing fiber bundle model and analyze the energy avalanche statistics of that model. We then compare the results with experiments and also present a plausible argument for its form.

\section{Description of Fiber Bundle Model}
After being introduced by Pierce in 1926 \cite{Pierce}, the fiber bundle model has been proven to be important yet arguably the simplest model to study failure processes in disordered solids. A conventional fiber bundle model consists of a set of linear elastic fibers or Hookean springs, attached between two parallel plates. The plates are pulled apart by a force $F$, creating a stress $\sigma=F/L$ on $L$ fibers. Once the stress crosses the breaking threshold of a particular fiber, chosen from a random distribution, that fiber breaks irreversibly. The stress of broken fibers is then redistributed either globally among all surviving fibers (global load sharing or GLS scheme) or among the surviving nearest neighbors only (local load sharing or LLS scheme). For the GLS scheme \cite{Pierce, Daniels} no stress concentration occurs anywhere around the failed fibers as the stress of the failed fibers is shared among all surviving fibers. On the other hand, in LLS scheme \cite{Phoenix,Smith,Newman,Harlow2,Harlow3,Smith2}, stress concentration is observed near a broken patch (set of neighboring broken fibers) and increases with the size of such patches. After such redistribution, the load per fiber increases initiating failure of more fibers starting an avalanche. At the end of an avalanche, either all fibers are broken (suggesting global failure) or the bundle comes to a stable state with few broken fibers where an increment of external stress is required to make the model evolve further. The last applied stress just before global failure is considered to be the nominal stress or strength $\sigma_c$ of the bundle. The fraction of fibers that survive at $\sigma_c$ just before global failure is defined as the critical unbroken fraction of fibers ($U_c$).

\begin{figure}[ht]
\centering
\includegraphics[width=8.7cm, keepaspectratio]{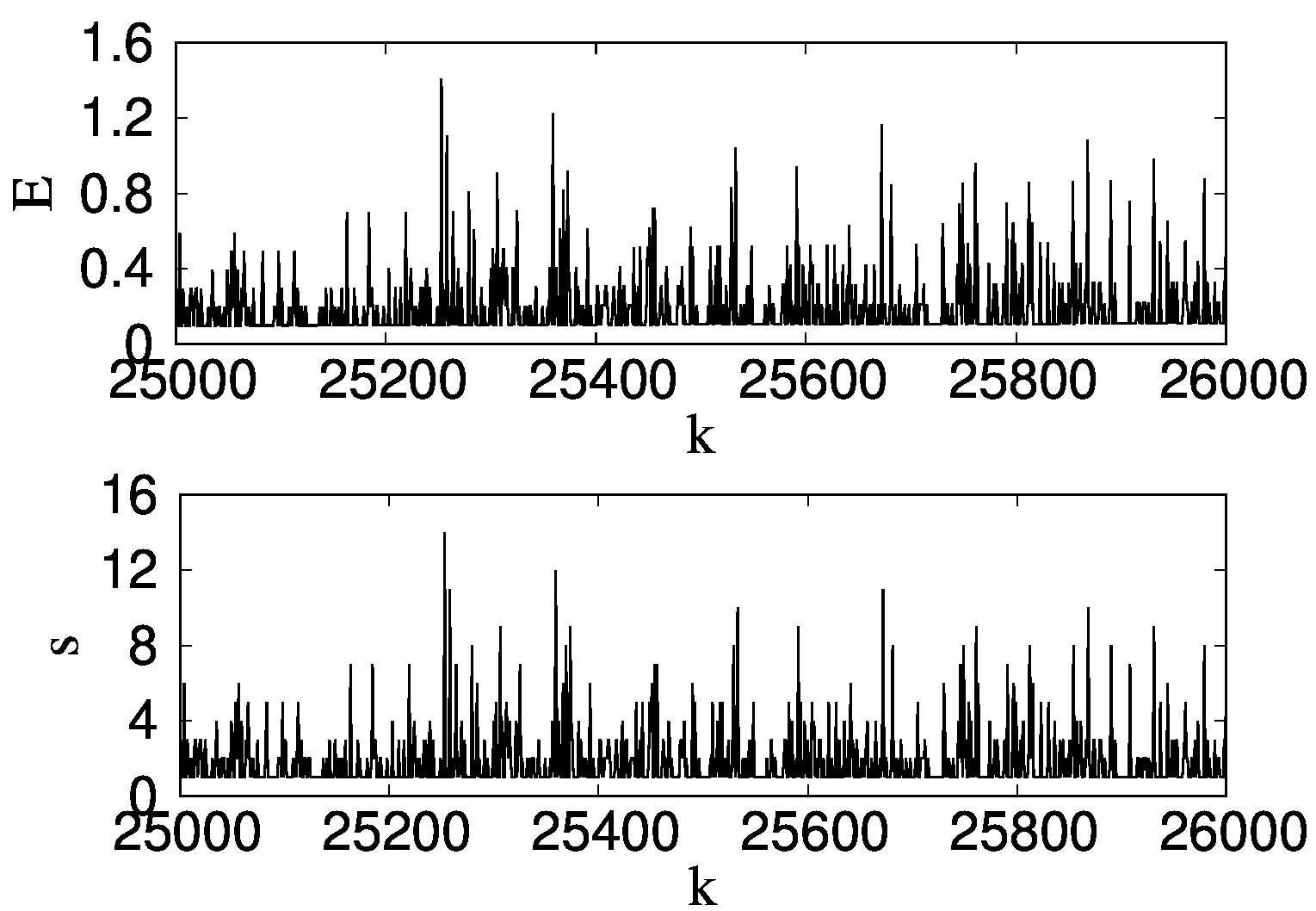} \includegraphics[width=8.5cm, keepaspectratio]{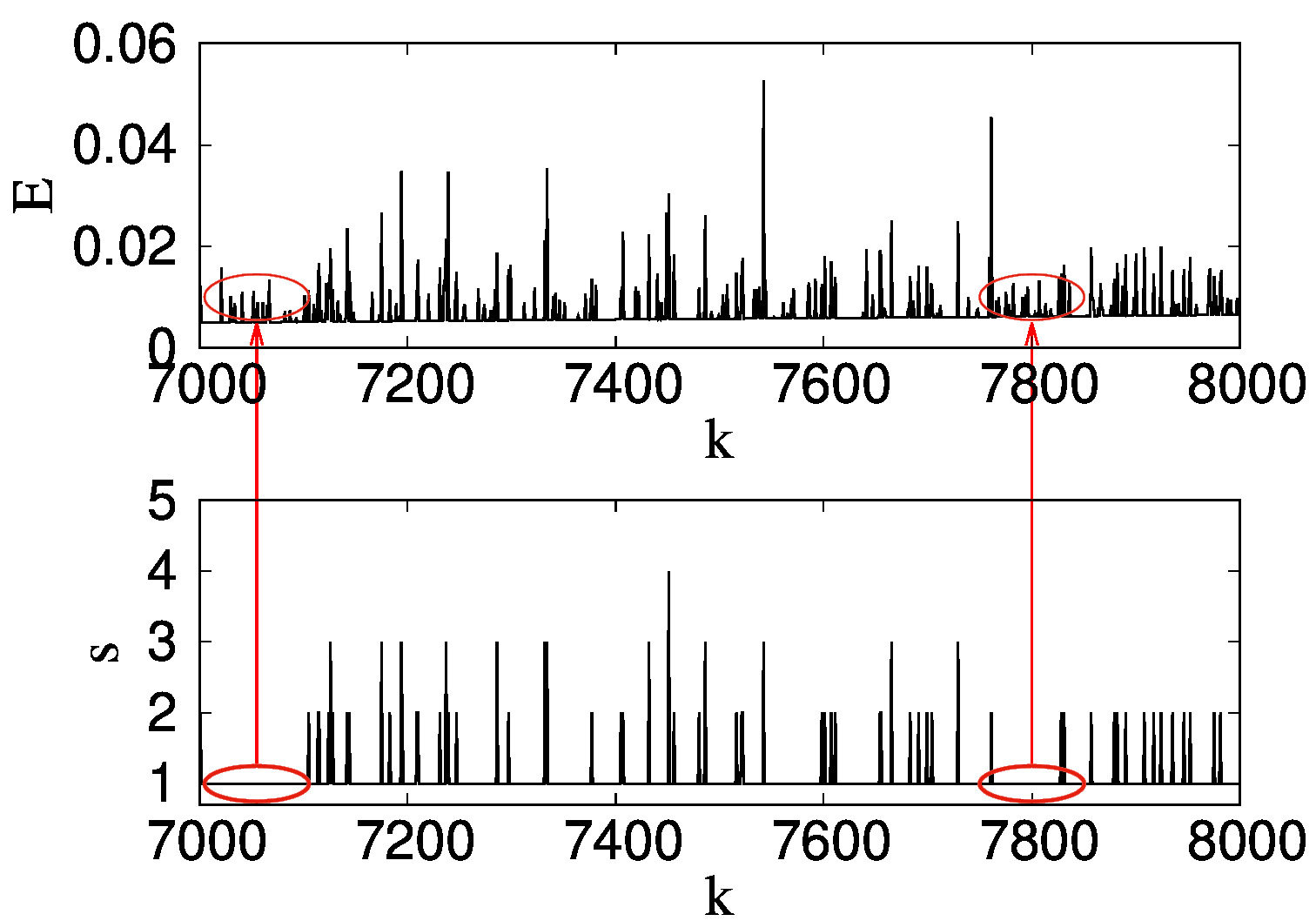}
\caption{Upper Panel: The figure shows the spectrum of avalanche sizes ($s$) and corresponding energy values ($E$) for GLS FBM with increasing number of avalanches $k$. We can see that there is a direct correspondence between the $s$ and $E$ values for a certain $k$. This means, higher $s$ gives higher $E$. Since in case of GLS FBM, the fibers break in an increasing order of threshold values, we get, $E(k+1)>E(k)$ if $s(k+1)=s(k)$.
Lower Panel: The figure shows the spectrum of avalanche sizes ($s$) and corresponding energy values ($E$) for LLS FBM with an increasing number of avalanches $k$. We do not see a direct correspondence between the $s$ and $E$ values here like GLS FBM. For example, the red eclipses show the parts where only 1 fiber breaks at each $k$ value but the corresponding $E$ values show many different values without any particular order as the fibers themselves do not follow any order while breaking.}
\label{fig1}
\end{figure}


\section{Numerical Results}

We have studied the fiber bundle model numerically in both mean-field limit and with local load sharing scheme in one dimension, though the major part of the paper will deal with the latter only. Numerical simulations are carried out for system sizes ranging in between $10^3$ and $5\times10^5$ and are averaged over $10^4$ configurations. Our motive is to understand the dynamics of avalanches and corresponding energy bursts emitted during these avalanches as the model evolves with increasing externally applied stress. Unless otherwise stated, we will use a uniform distribution ranging from 0 to 1 in order to assign threshold values to individual fibers beyond which it breaks. 


\subsection{Relation between $s$ and $E$}

Figure \ref{fig1} shows a comparison between different avalanches and energy emitted during those avalanches for a bundle of size $10^5$. The results are produced for a single configuration. As usual, an avalanche is defined as the number of fibers broken in-between two consecutive stress increments; $k$ is the number of such stress increments in this case. While presenting the energy spectrum and the avalanches we have excluded the final avalanche leading to global failure.

\begin{figure}[ht]
\centering
\includegraphics[width=8.5cm, keepaspectratio]{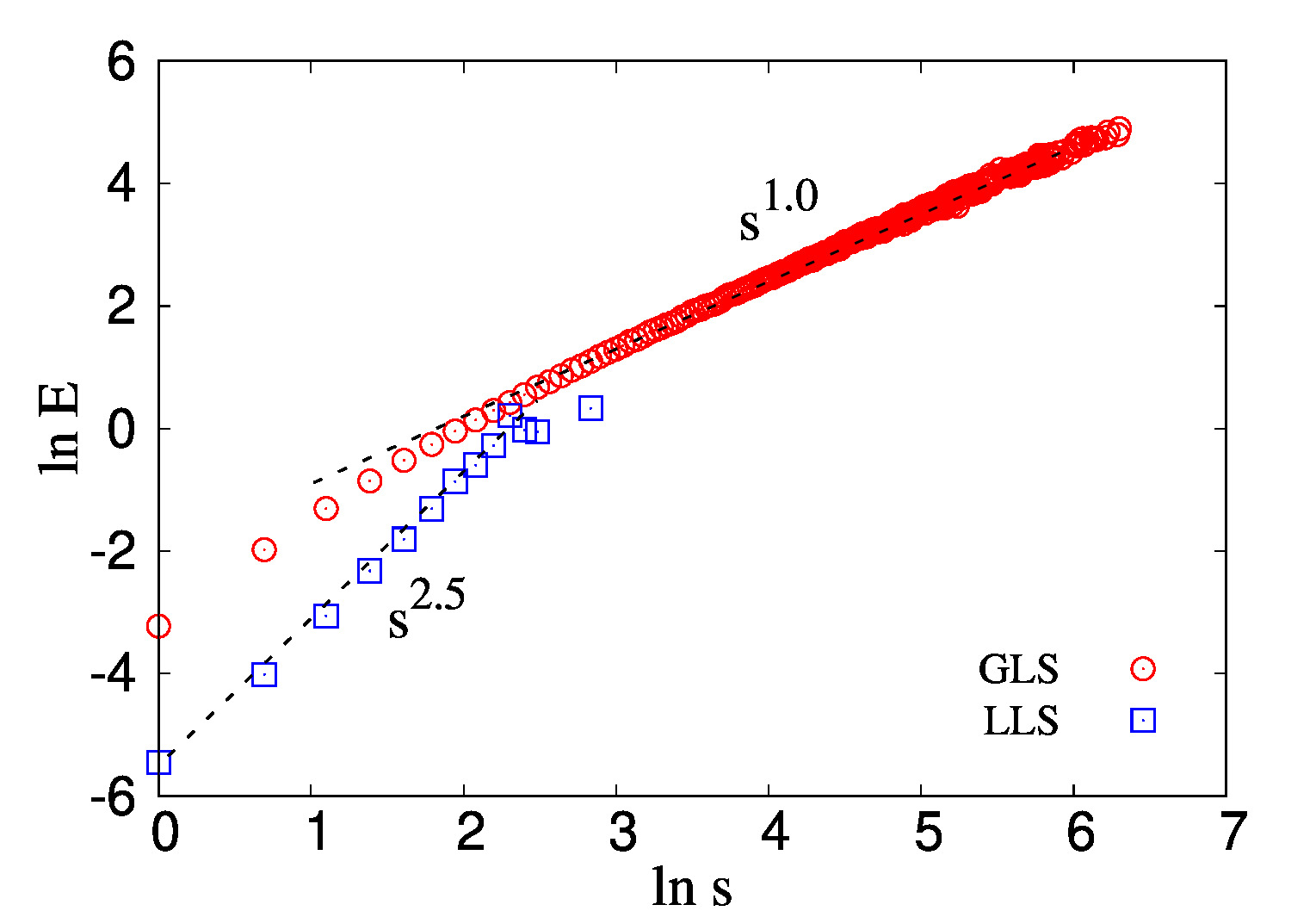}
\caption{The figure shows the variation of energy $E$ with avalanche size (total number of broken fibers) $s$ for both GLS and LLS fiber bundle model. We observe $E \sim s$, for GLS FBM. On the other hand, for LLS scheme, $E \sim s^{\gamma}$, with $\gamma \approx 2.5$.}
\label{fig2}
\end{figure}

The upper panel of figure \ref{fig1} shows the results for the GLS fiber bundle model while on the lower panel, we have shown the results with the local load sharing (LLS) scheme. Note that the range of $k$ for the LLS model is much less than the range of $k$ with the GLS scheme. This is understandable since with the LLS scheme, the model is more unstable due to stress concentration and a large number of fibers are broken during the final avalanche. The model evolves with a lesser number of stress increments in this case prior to a global failure where the average size of the avalanches is smaller compared to that in the GLS scheme. Now, for an avalanche of size $s$, if $n$ fibers with threshold values $\tau_1$, $\tau_2$, $\tau_3$, $\cdots$, $\tau_n$ break, then the amount of energy emitted during this avalanche will be:
\begin{align}\label{eq1}
E(s) = \displaystyle\frac{1}{2}\displaystyle\sum_{i=1}^{n} \tau_i^2. 
\end{align}   
This follows from the assumption of linear elastic (stress $\propto$ strain) behavior of individual fibers up until their individual (brittle) failure points.
With above formalism, for each stress increment $k$, we will obtain an avalanche $s(k)$ and a corresponding energy burst of magnitude $E(s(k))$. \\

The energy spectrum follows a particular trend in the case of the GLS scheme. Since with the GLS scheme the fibers break in the increasing order of their threshold values, the energy emitted at $k+1$-th load increment will be higher than the energy emitted at $k$-th increment, even if the avalanche sizes happen to be the same at $k$ and $k+1$. Due to this, the variations of $s$ and $E$ with increasing $k$ looks exactly the same, only the values are scaled by a constant when we transfer from $s$ to $E$. Such correlation between $s$ and $E$ is not present in the case of the LLS fiber bundle model. In the case of the LLS scheme, the fibers break due to the interplay between the local stress profile and the threshold values of the fibers themselves. Due to such dynamics, the fibers do not break in increasing order of their thresholds. Then, there might be scenarios where $E(k+1)<E(k)$ when $s(k+1)=s(k)$ or even $s(k+1)>s(k)$. The red ellipses in the lower panel of figure \ref{fig1} shows this absence of correlation between $s(k)$ and $E(s(k))$. For both ellipses, $s=1$ for that period. In spite of that, we see a fluctuation in energy values without a particular trend. In the following, we will discuss this relation between $s$ and $E$ in detail. 

\begin{figure}[ht]
\centering
\includegraphics[width=8cm, keepaspectratio]{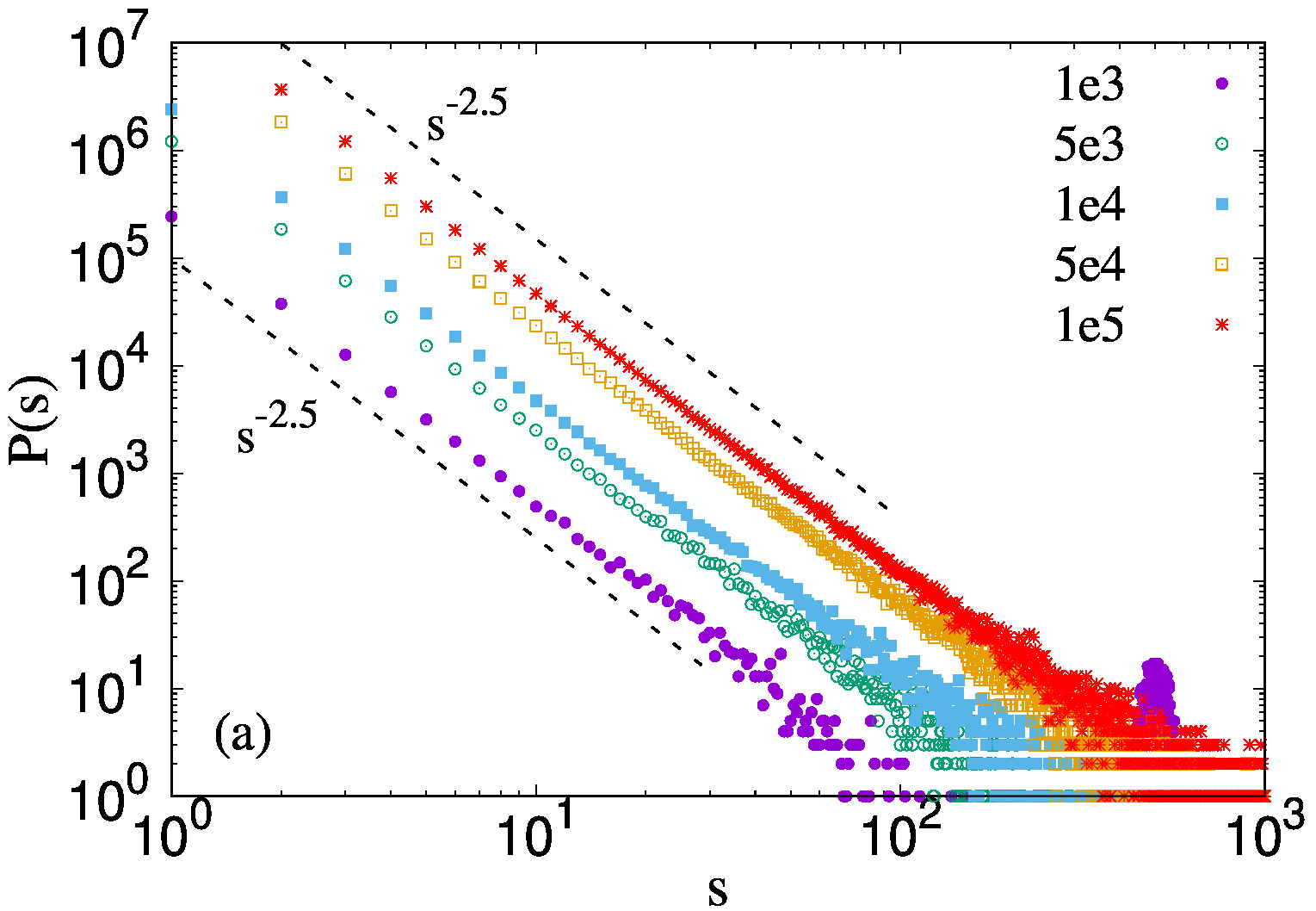} \\ \includegraphics[width=8cm, keepaspectratio]{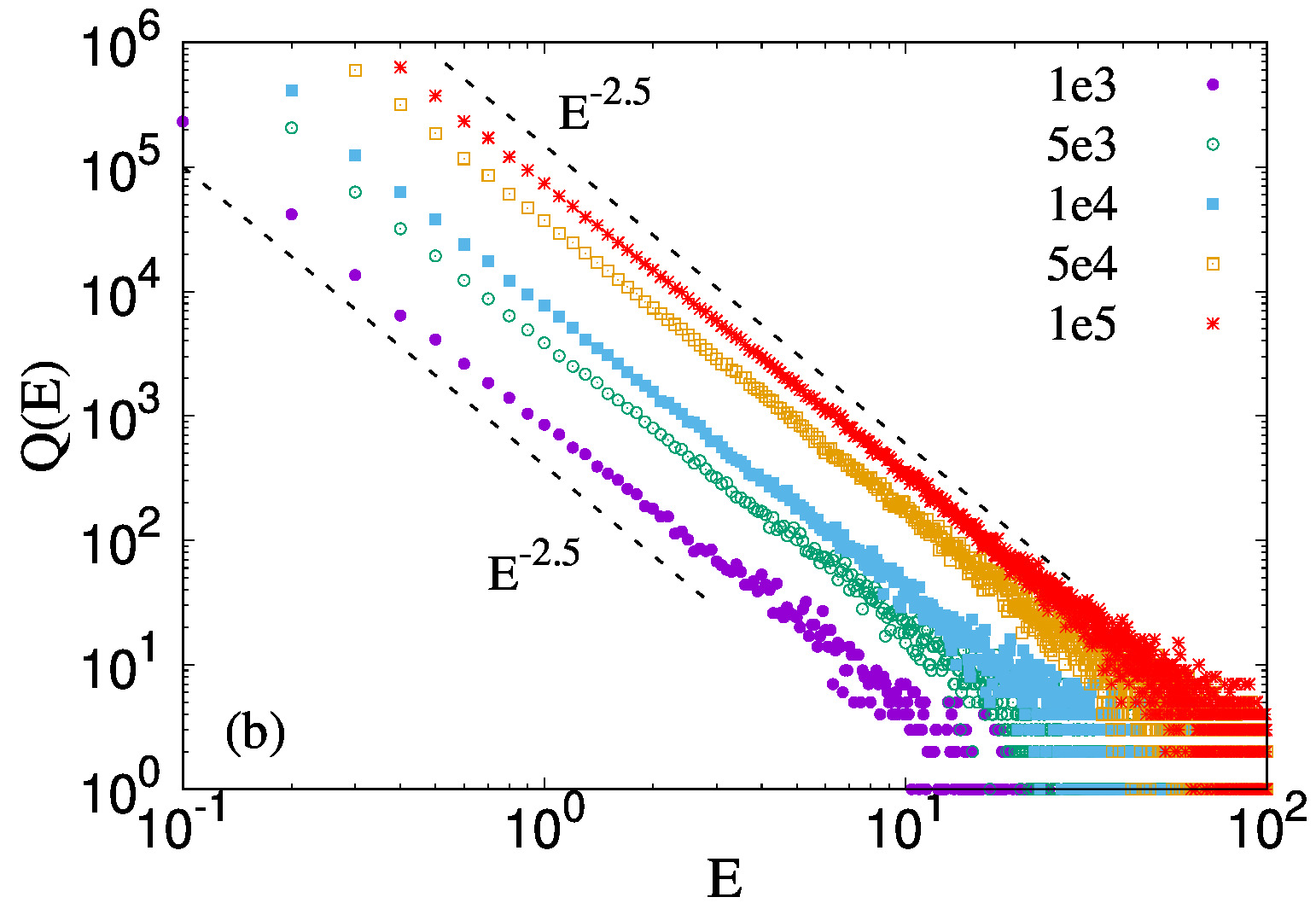}
\caption{Distribution of energies for an uniform distribution (0,1) and system sizes ranging in between $10^3$ and $10^5$. The results are shown for GLS FBM. (a) We already know that in the mean-field limit $P(s) \sim s^{-\beta}$, with $\beta \approx 2.5$. (b) We observe $Q(E) \sim E^{-\alpha}$ where $\alpha \approx 2.5$ as well independent of the system size.}
\label{fig3}
\end{figure}

Figure \ref{fig2} highlights how energy $E$ emitted as a function of average avalanche size $s$ for a bundle of size $10^5$ and configuration $10^4$. Results for both GLS and LLS schemes are shown in the figure. We observe the following behavior:
\begin{equation}
\label{eq2}
E \sim \left\{\begin{array}{ll}
                                 s       & \mbox{, \text{for GLS},}\\
                                 s^{\gamma} & \mbox{, \text{with $\gamma=2.5$ for LLS}.}\\
                  \end{array}  
                  \right.
\end{equation}  
This behavior can be used to understand the relation between distributions $P(s)$ of avalanche size $s$ and $Q(E)$ of emitted energies $E$. For this we simply need to implement a change in variable \cite{theorem1} scheme as follows:
\begin{align}\label{eq3}
Q(E) \sim P[s(E)].|s^{\prime}(E)| = P[s(E)].|\displaystyle\frac{ds(E)}{dE}|
\end{align}  

\textbf{Change in variable: GLS scheme}

In case of GLS scheme, we observe 
\begin{align}\label{eq4}
&E(s) \sim s \nonumber \\
&s(E) \sim E
\end{align}
This makes 
\begin{align}\label{eq5}
s^{\prime}(E) = \displaystyle\frac{ds(E)}{dE} \sim 1
\end{align}
We also know that the avalanche size distribution in case of GLS scheme is a scale free distribution with an exponent 2.5 \cite{hh92}.
\begin{align}\label{eq6}
P(s) \sim s^{-\beta}, \ \ \ \text{with $\beta=2.5$}
\end{align}
Then, combining Eq.\ref{eq3}, \ref{eq4}, \ref{eq5} and \ref{eq6}, we get,
\begin{align}\label{eq6a}
Q(E) \sim P(E).1 \sim E^{-\beta}
\end{align}

\begin{figure}[ht]
\centering
\includegraphics[width=8cm, keepaspectratio]{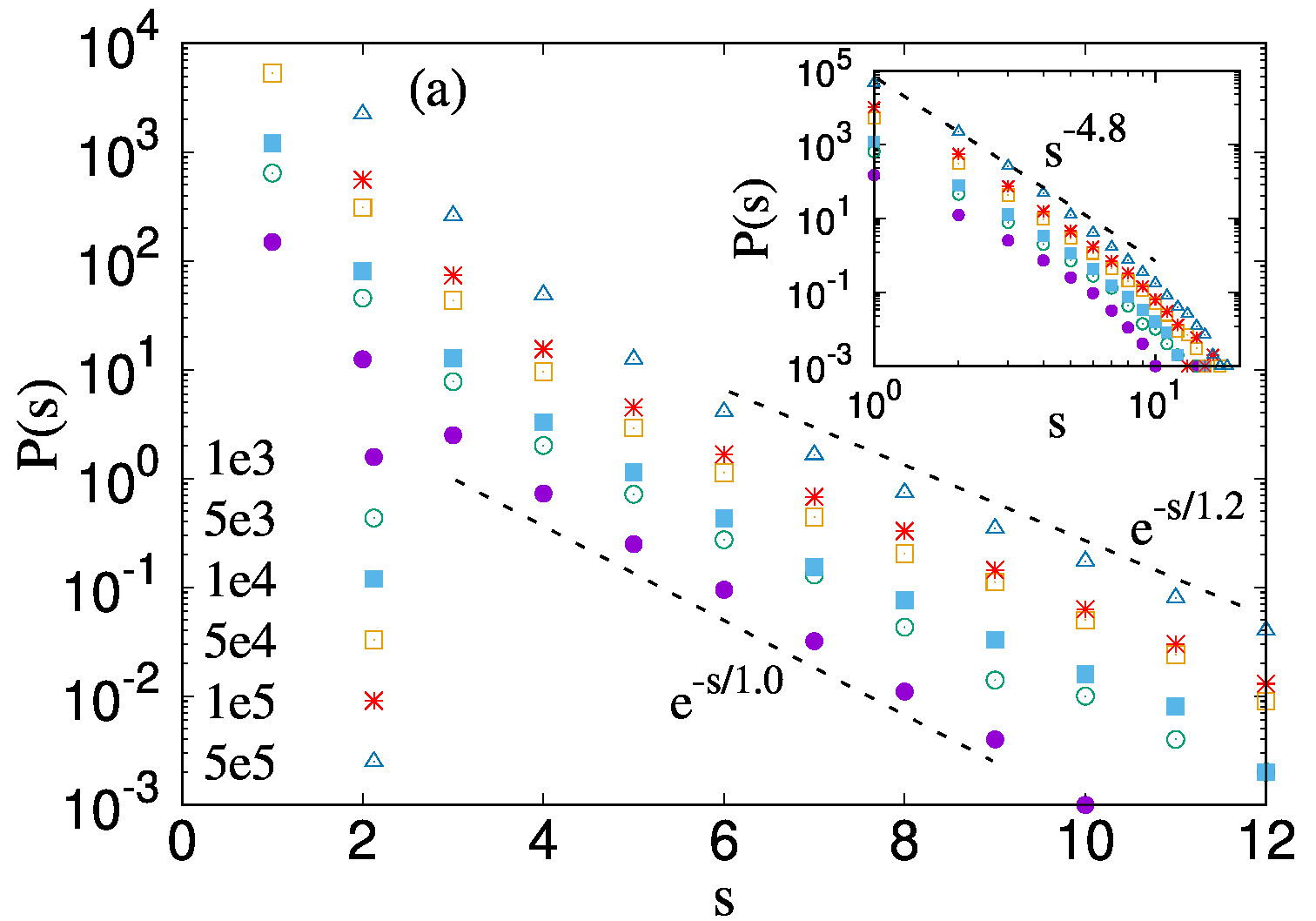} \\ \includegraphics[width=8cm, keepaspectratio]{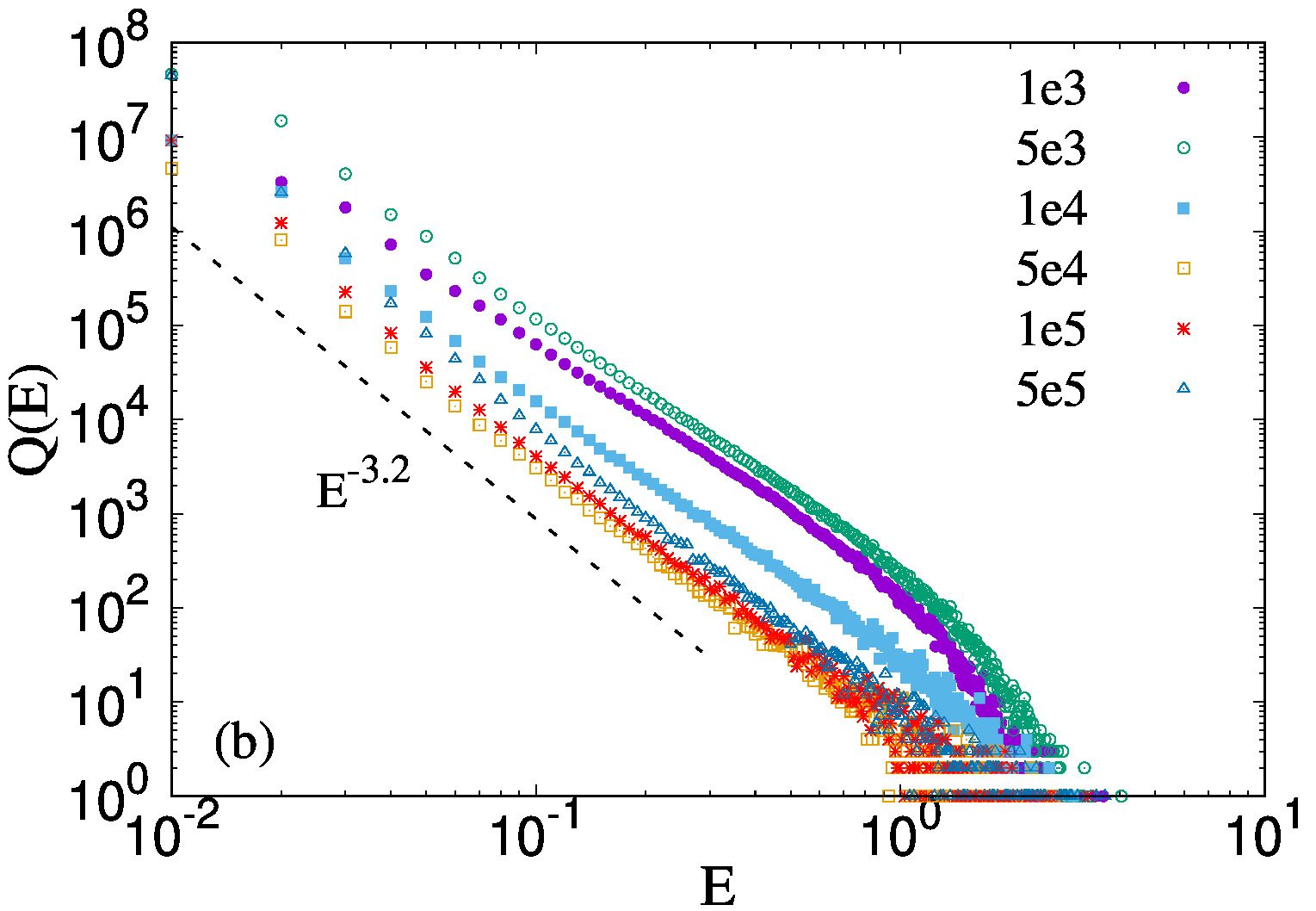}
\caption{Distribution of energies for an uniform distribution [0:1] and system sizes ranging in between $10^3$ and $10^5$. The results are shown for LLS FBM. (a) Avalanche size distribution for LLS FBM is an exponential function: $P(s) \sim e^{-s/s_0}$, where $s_0$ depends weakly on the system size. (b) Scale free distribution for energy emitted: $Q(E) \sim E^{-\alpha}$, with $\alpha \approx 3.5$.}
\label{fig4}
\end{figure}
\begin{figure*}[ht]
\centering
\includegraphics[width=15cm, keepaspectratio]{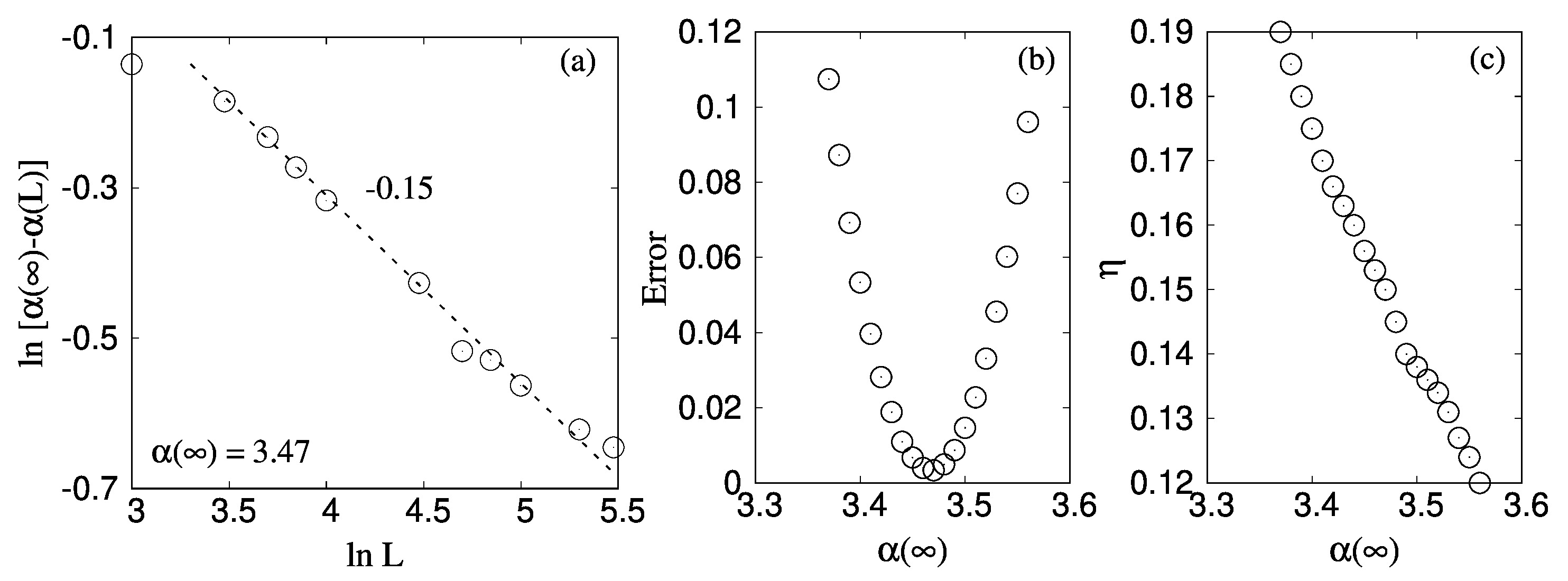}
\caption{(a) The finite size effect of $\alpha(L)$ is shown, which follows a scaling: $[\alpha(\infty)-\alpha(L)] \sim L^{-\eta}$ with $\eta=0.15$, where $\alpha(\infty)$ is the value of the exponent in the thermodynamic limit. We get $\alpha(\infty)=3.47$.  (b) \& (c) The least square fit error and corresponding exponent $\eta$ for the scaling $[\alpha(\infty)-\alpha(L)] \sim L^{-\eta}$ is given for different values of $\alpha(\infty)$. We consider the value of $\alpha(\infty)$ and $\eta$ which produces the minimum error. This same procedure has been followed next while exploring the same thing for different threshold distributions.}
\label{fig5}
\end{figure*}

\textbf{Change in variable: LLS scheme}

In case of LLS scheme, we observe 
\begin{align}\label{eq7}
&E(s) \sim s^{\gamma} \nonumber \\
&s(E) \sim E^{-\gamma}
\end{align}
This makes 
\begin{align}\label{eq8}
s^{\prime}(E) = \displaystyle\frac{ds(E)}{dE} \sim (-\gamma)E^{-(\gamma+1)}
\end{align}
where $\gamma=2.5$. We also know that the avalanche size distribution in case of LLS scheme is an exponential distribution \cite{khh97}.
\begin{align}\label{eq9}
P(s) \sim e^{-s/s_0}
\end{align}
Then, combining Eq.\ref{eq3}, \ref{eq7}, \ref{eq8} and \ref{eq9}, we get,
\begin{align}\label{eq10}
Q(E) \sim P(E).\gamma E^{-(\gamma+1)} \sim \gamma e^{-\displaystyle\frac{E^{-\gamma}}{s_0}} E^{-(\gamma+1)}
\end{align}
In the limit of high $E$ value, Eq.\ref{eq10} can be simplified as follows
\begin{align}\label{eq11}
Q(E) \sim E^{-\alpha} \ \ \ \text{where $\alpha=\gamma+1=3.5$}
\end{align}
Above treatment shows that, in case of LLS scheme, in spite of an exponential distribution for avalanche sizes, the distribution of emitted energy is still observed to be scale-free. 


\subsection{Distribution of $s$ and $E$: Uniform distribution}

Figure \ref{fig4}(a) shows the avalanche size distribution $P(s)$ for a GLS fiber bundle model with system size ranging from $10^3$ to $10^5$. This scale-free decrease of $P(s)$ with $s$ is already known in the literature. We also observe the same universal exponent 2.5 \cite{hh92}. Figure \ref{fig4}(b) shows the corresponding distribution for the energy emitted. We observe the same scale-free distribution for the energy as well, with the same exponent 2.5. This behavior is consistent with Eq.\ref{eq6} and Eq.\ref{eq6a} respectively.

Figure \ref{fig4}(a), on the other hand, shows the avalanche size distribution with the LLS scheme. The distribution is exponential as derived analytically by Kloster et al.\cite{khh97}. The inset of the same results in log scale in order to compare them with the previous claim by Zhang and Ding \cite{zd94}, that $P(s)$ shows a scale-free behavior with a very high exponent closer to $-4.8$. This claim of scale-free nature is not substantiated and the exponential form for $P(s)$ is accepted in the literature.     

The distribution of energy in figure \ref{fig4}(b) shows a scale-free distribution, in spite of the fact that the avalanche size distribution is an exponential distribution. The exponent of the scale-free distribution is observed to an increasing function of the size of the bundle
\begin{align}\label{eq12}
Q(E) \sim E^{-\alpha(L)}
\end{align}
The above behavior is similar to Eq.\ref{eq11}, but with a $L$ dependent exponent instead of a constant value. To compare this $L$ dependent exponent with the value in Eq.\ref{eq11}, we have to study the variation of $\alpha$ in Eq.\ref{eq12} in details as the size of the bundle is increased. We have discussed this next.

Figure \ref{fig5} shows the scaling of exponent $\alpha$ in figure \ref{fig4}(b) as the model approaches the thermodynamic limit. We observe the following scaling, 
\begin{align}\label{eq13}
\alpha(\infty)-\alpha(L) \sim L^{-\eta}
\end{align}
where $\eta=-0.15$ and $\alpha(\infty)$ (= 3.47) has a value close to $\gamma+1$ (see Eq.\ref{eq11}). The fitting and the exponent $\eta$ is calculated from the minimization of least square fit error. This is shown in figure \ref{fig5}(b) and (c). We choose a certain value of $\alpha(\infty)$ and fit our numerical results. This in turn will produce a value of $\eta$ and corresponding least square fit error. If we repeat this for a number of $\alpha(\infty)$ values, then we can express the error (see figure \ref{fig5}b) and the exponent $\eta$ (see figure \ref{fig5}c) as a function of $\alpha(\infty)$. The dotted line in figure \ref{fig5}(a) corresponds to the value of $\alpha(\infty)$ (= 3.47) and $\eta$ (= 0.15) for which the least square fit error is minimum.


\subsection{Universality}

So far, we have generated the numerical results where a uniform distribution from 0 to 1 is used to assign random thresholds to individual fibers. In this section, we will verify the universality of our results. For this purpose, we will mainly explore 4 other distributions: (i) linearly increasing from 0 to 1, (ii) linearly decreasing from 0 to 1, (ii) a Weibull distribution with scale parameter 1 and Weibull modulus 1, and (iii) A power law distribution from 0 to 1 with exponent 2.0.

\begin{figure*}[ht]
\centering
\includegraphics[width=8cm, keepaspectratio]{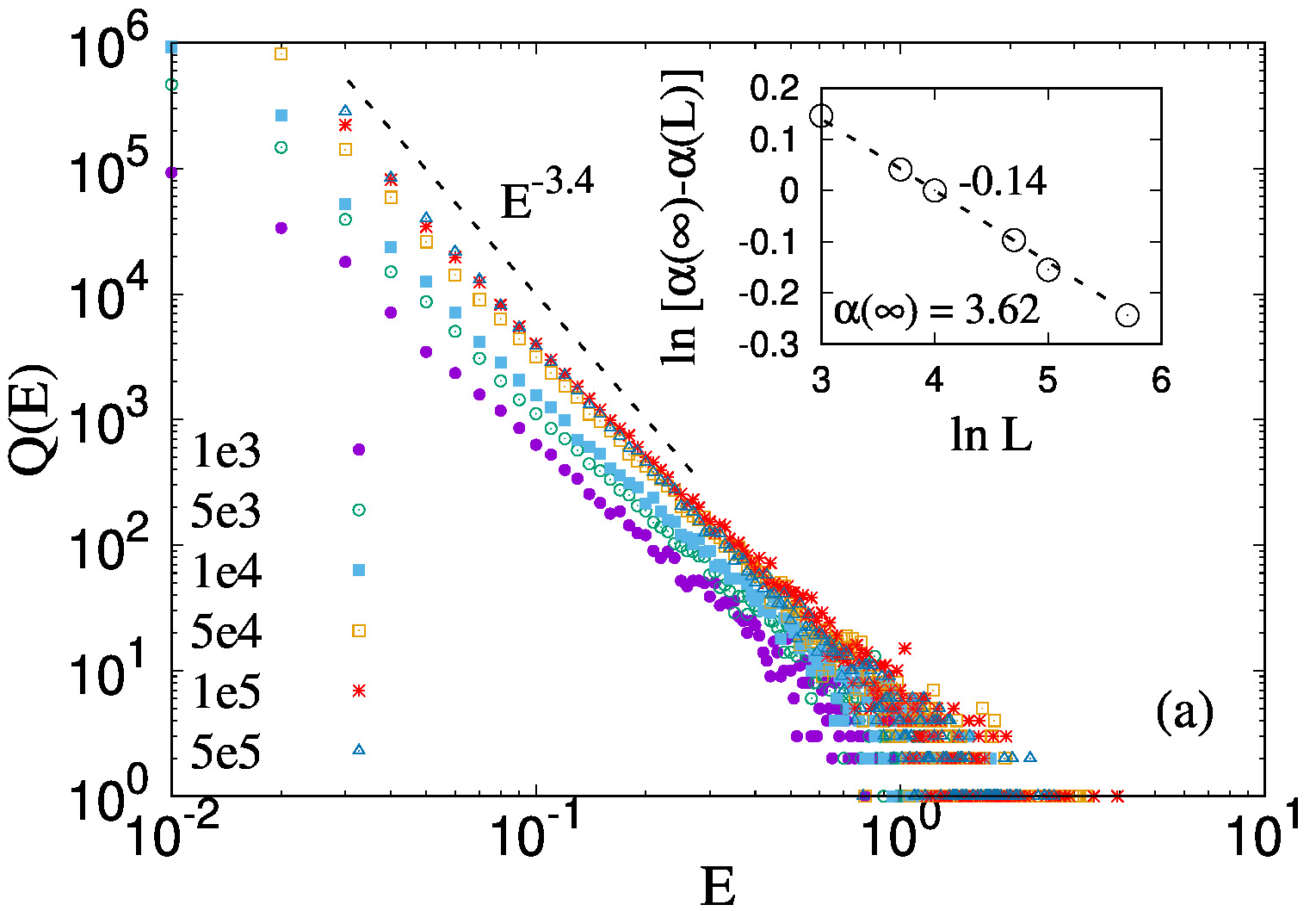} \ \ \ \includegraphics[width=8cm, keepaspectratio]{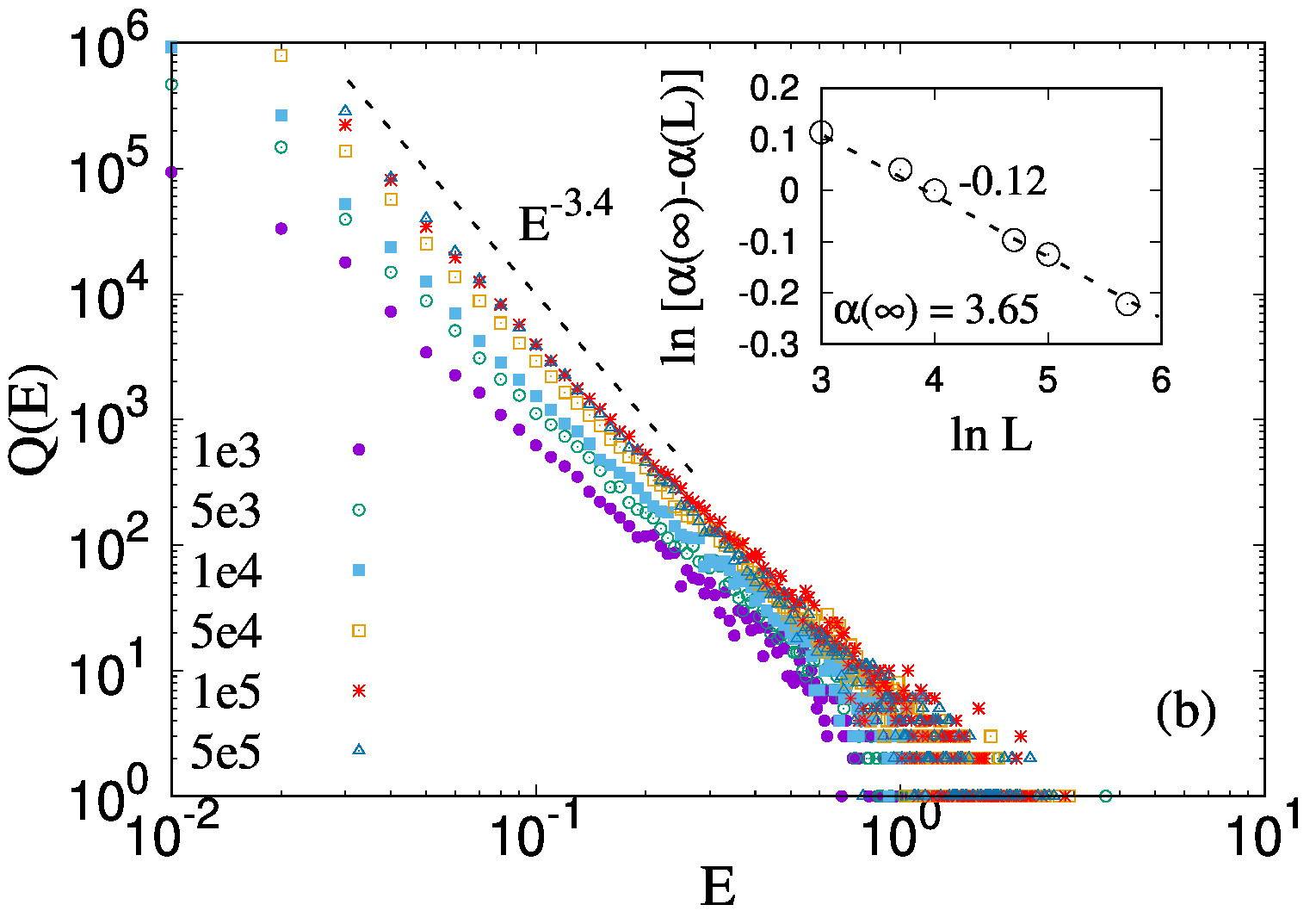} \\ \includegraphics[width=8cm, keepaspectratio]{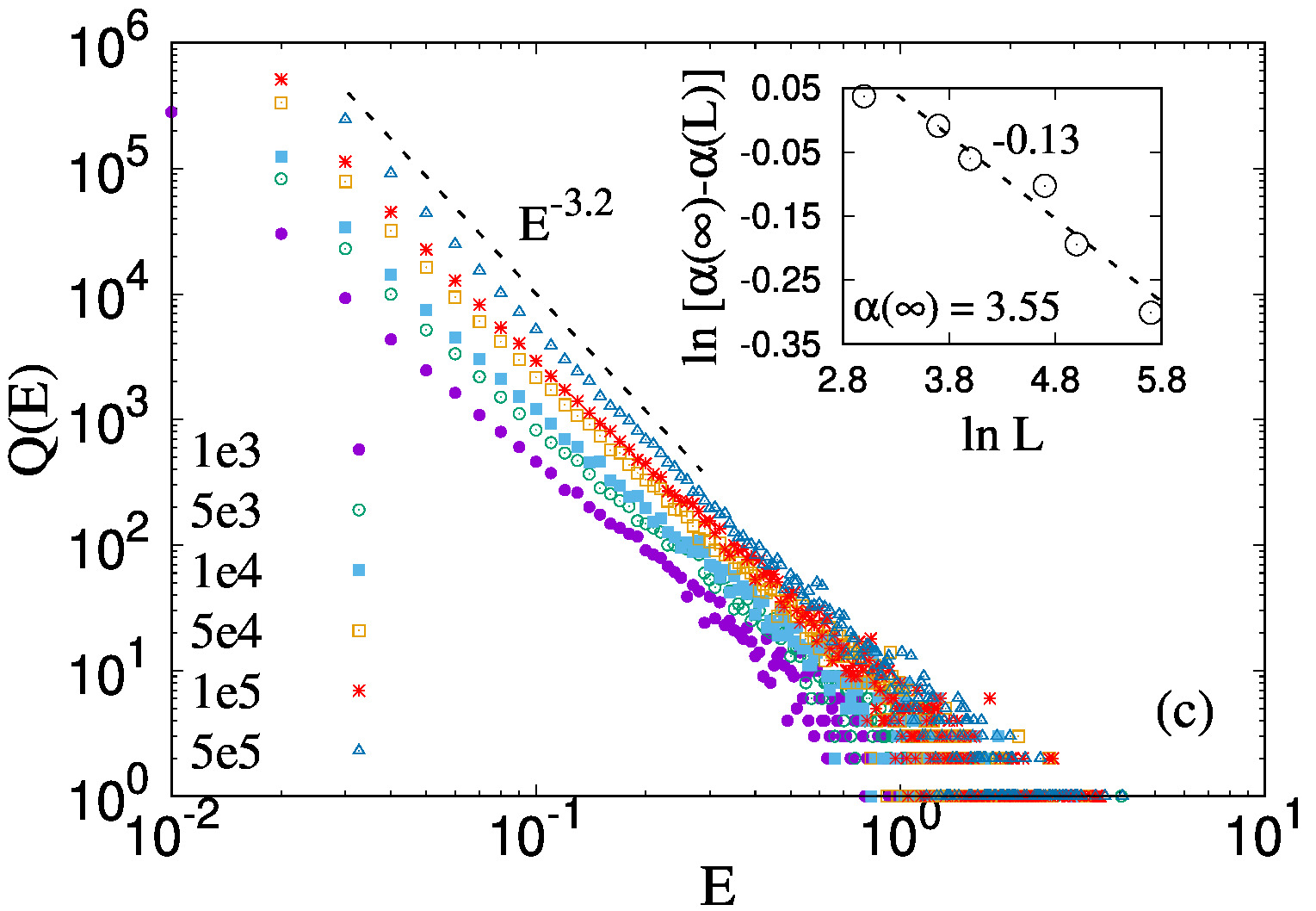} \ \ \ \includegraphics[width=8cm, keepaspectratio]{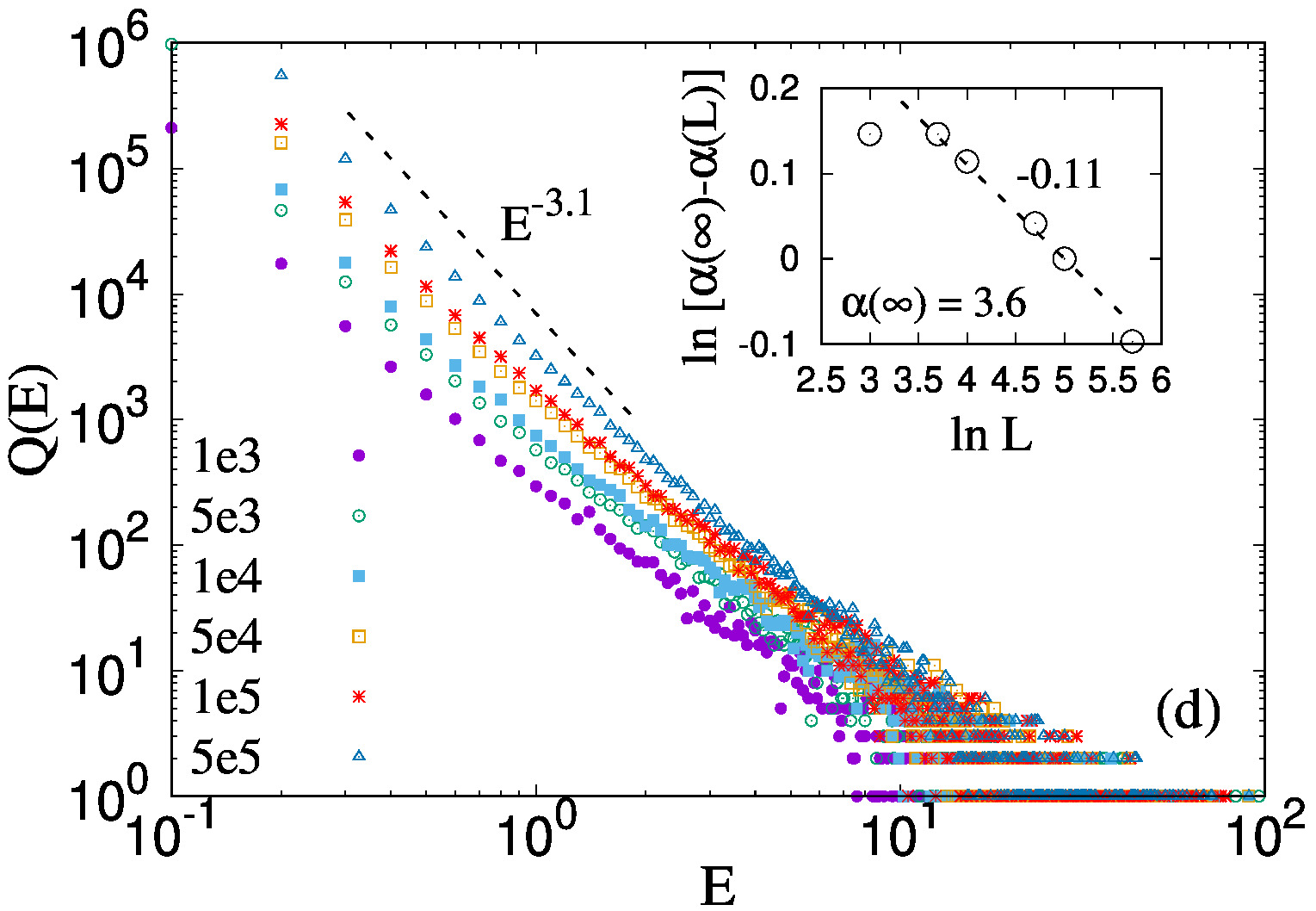}
\caption{(a) Distribution of energies for 4 different threshold distributions: (a) Linearly increasing [0:1], (b) Linearly decreasing [0:1] (c) Scale free distribution of exponent 2 between 0 and 1, (d) Weibull distribution with scale factor 1.0 and Weibull modulus 1.0. The system sizes varies in between $10^3$ and $5\times10^5$. The results are shown for LLS FBM. We observe a scale free distribution for $E$: $Q(E) \sim E^{-\alpha}$ for all thresholds. The exponent value $\alpha(L)$ shows finite size effect. The dotted line shows the slope with highest system size in our simulation. The exponents are close but less than 3.5. As before, this obeys the following scaling: $[\alpha(\infty)-\alpha(L)] \sim L^{-\eta}$. The value of $\eta$ for above mentioned distributions are 0.14, 0.12, 0.13 and 0.11 respectively. We observe the exponent $\alpha(\infty)$ in the thermodynamic limit to be very close to 3.5 irrespective of the choice of the threshold distribution.}
\label{fig6}
\end{figure*}
\begin{figure}[ht]
\centering
\includegraphics[width=8cm, keepaspectratio]{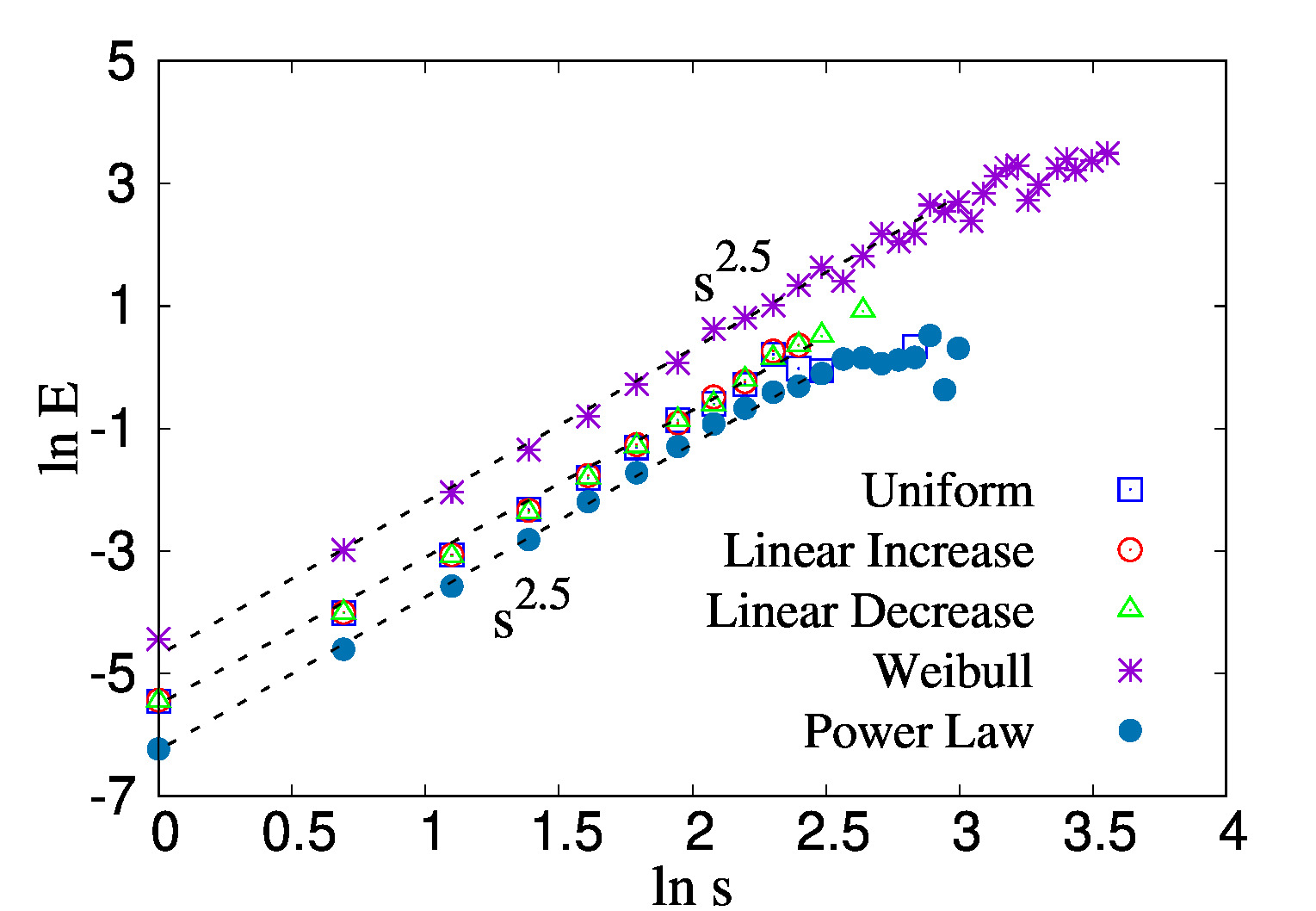}
\caption{The figure shows the variation of energy $E$ with avalanche size (total number of broken fibers) $s$ for LLS fiber bundle model. We have repeated the study for 5 different threshold distributions. We observe for all distributions, $E \sim s^{\gamma}$, where $\gamma$ has a value 2.5 independent of the nature of the distribution.}
\label{fig7}
\end{figure}

In all these cases, the energy burst size distributions were found to be scale-free with an exponent value close to $-3.5$ (see figure \ref{fig6}), as is predicted from Eq. (\ref{eq11}). The variation with system size also universal across these different threshold distributions. These results suggest that the scale-free nature of the energy burst size distribution in the local load sharing fiber bundle model is a universal feature. 

We have further checked that the relation between an avalanche size and an energy burst size i..e., $E\sim s^{2.5}$ is valid for all these threshold distributions, as can be seen from Fig. \ref{fig7}.


\section{Discussions and conclusions}
The local load sharing fiber bundle model is known to be lacking in reproducing the scale-free avalanche statistics often seen in the experimental setup of fracturing brittle solids. In all the interpolation schemes between global (equal) and local load sharing versions of fiber bundles, the avalanche size distribution $P(s)$ only show a cross-over between the mean-field ($P(s)\sim s^{-\beta}$) and local load sharing ($P(s)\sim e^{-s/s_0}$) limits. The mean-field limit, however, is a rather idealized condition for modeling real samples.

However, one important distinction between avalanche sizes ($s$) of the fiber bundle model and what is usually measured in the experiments is that in the latter case it is the energy burst ($E$) emitted in an avalanche that is measured. However, that distinction is not at all significant in the mean-field i.e., the global load-sharing limit of the model, because in that limit $E\sim s$. However, in the local load sharing version, we numerically find $E\sim s^{\gamma}$. Given an exponential distribution for the avalanche size distribution in the local load sharing limit and this numerical observation, it is possible to show that the size distribution of the energy bursts is scale-free ($Q(E)\sim E^{-\alpha}$) with $\alpha=\gamma+1$ (see Eq. (\ref{eq11})). We have then numerically checked that $\gamma\approx 2.5$ for various different threshold distributions (see Fig. \ref{fig6}) and independently checked that the size distribution exponent for the energy bursts are close to $-3.5$ (see Figs. \ref{fig4}, \ref{fig7}). Indeed, there are indications in experiments with sandstones that the avalanche amplitude distribution was exponential while the energy burst distribution was found to be a power law (see e.g., \cite{arma,expt}). Our results reproduce the same for the local load sharing fiber bundle model.    

In conclusion, the local load sharing fiber bundle model is shown to have a non-trivial relation between the avalanche size (number of fibers broken) and the energy burst size (elastic energy released from the broken fibers). Consequently, the energy burst size distribution is shown to have scale-free nature, with an exponent value independent of the threshold distributions of the fibers. Given that experimentally one measures the energy released, these results indicate that local load sharing fiber bundles can have a significant role in modeling fracture of brittle solids without having to resort to the equal load sharing mean-field limit. 


\bigskip

The authors thank Prof. Bikas K. Chakrabarti for his valuable comments. This work was partly supported by the Research Council of Norway through its Centres of Excellence funding scheme, project number 262644.




\begin{thebibliography}{99}
\bibitem{theorem1} Let's assume $x$ and $y$ are a continuous functions with p.d.f $f(x)$ and $g(y)$. Also, $y=u(x)$ is a function with inverse, which means it is possible to find $x=v(y)$. In this circumstances, $g(y)$ and $f(x)$ can be expressed as: $g(y)=f[v(y)].|v^{\prime}(y)|$, where $|v^{\prime}(y)|=|dv(y)/dy|$.
\end{thebibliography}
\end{document}